\documentclass[aps,prl,twocolumn,nopacs,superscriptaddress,longbibliography]{revtex4}
\usepackage[breaklinks=true,colorlinks,citecolor=blue,linkcolor=blue,urlcolor=blue]{hyperref}
\usepackage{epsfig,mathrsfs,color,latexsym,subfigure,marginnote,graphicx,verbatim,relsize,mathrsfs,color,array,amsmath,amsfonts,amssymb,graphicx}
\renewcommand{\BibitemShut}[1]{}

\newcommand\scalemath[2]{\scalebox{#1}{\mbox{\ensuremath{\displaystyle #2}}}}

\begin{document}
\title{Prediction of three-fold fermions in a nearly-ideal Dirac semimetal BaAgAs}

\author{Sougata Mardanya$^*$}
\affiliation{Department of Physics, Indian Institute of Technology Kanpur, Kanpur 208016, India}

\author{Bahadur Singh\footnote{These authors contributed equally to this work.}$^{\dag}$}
\affiliation{SZU-NUS Collaborative Center and International Collaborative Laboratory of 2D Materials for Optoelectronic Science $\&$ Technology, 
Engineering Technology Research Center for 2D Materials Information Functional Devices and Systems of Guangdong Province, College of Optoelectronic Engineering, Shenzhen University, Shenzhen 518060, China}
\affiliation{Department of Physics, Northeastern University, Boston, Massachusetts 02115, USA}

\author{Shin-Ming Huang}
\affiliation{Department of Physics, National Sun Yat-sen University, Kaohsiung 80424, Taiwan}
\affiliation{Center of Crystal Research, National Sun Yat-sen University, Kaohsiung, 80424, Taiwan}

\author{Tay-Rong Chang}
\affiliation{Department of Physics, National Cheng Kung University, Tainan 701, Taiwan}
\affiliation{Center for Quantum Frontiers of Research $\&$ Technology (QFort), National Cheng Kung University, Tainan 701, Taiwan}

\author{Chenliang Su}
\affiliation{SZU-NUS Collaborative Center and International Collaborative Laboratory of 2D Materials for Optoelectronic Science $\&$ Technology, 
Engineering Technology Research Center for 2D Materials Information Functional Devices and Systems of Guangdong Province, College of Optoelectronic Engineering, Shenzhen University, Shenzhen 518060, China}

\author{Hsin Lin$^{\dag}$}
\affiliation{Institute of Physics, Academia Sinica, Taipei 11529, Taiwan}

\author{Amit Agarwal$^{\dag}$}
\affiliation{Department of Physics, Indian Institute of Technology Kanpur, Kanpur 208016, India}

\author{Arun Bansil\footnote{Corresponding authors' emails: bahadursingh24@gmail.com, nilnish@gmail.com, amitag@iitk.ac.in, ar.bansil@northeastern.edu}}
\affiliation{Department of Physics, Northeastern University, Boston, Massachusetts 02115, USA}

\begin{abstract}
Materials with triply-degenerate nodal points in their low-energy electronic spectrum produce crystalline-symmetry-enforced three-fold fermions, which conceptually lie between the two-fold Weyl and four-fold Dirac fermions. Here we show how a silver-based Dirac semimetal BaAgAs realizes three-fold fermions through our first-principles calculations combined with a low-energy effective $\mathbf{k.p}$ model Hamiltonian analysis. BaAgAs is shown to harbor triply-degenerate nodal points, which lie on its $C_{3}$ rotation axis, and are protected by the $C_{6v}$($C_2\otimes C_{3v}$) point-group symmetry in the absence of spin-orbit coupling (SOC) effects.  When the SOC is turned on, BaAgAs transitions into a nearly-ideal Dirac semimetal state with a pair of Dirac nodes lying on the $C_{3}$ rotation axis. We show that breaking inversion symmetry in the BaAgAs$_{1-x}$P$_x$ alloy yields a clean and tunable three-fold fermion semimetal. Systematic relaxation of other symmetries in BaAgAs generates a series of other topological phases. BaAgAs materials thus provide an ideal platform for exploring tunable topological properties associated with a variety of different fermionic excitations.
\end{abstract}
\maketitle

\textcolor{blue}{$ {\it Introduction}.$\textemdash} 
Topological semimetals are currently drawing intense interest in condensed matter and materials physics\cite{Bansil2016,Ashvin2018,kong11}. In addition to their potential use as platforms for next-generation electronics/spintronics device applications, they provide a fertile ground for exploring relativistic particles and high-energy phenomenology at the far more accessible solid-state physics scale. Well-known examples are Dirac and Weyl semimetals in which electronic states near the band crossings mimic the Dirac\cite{young12,wang12,liu14a,liu14b,neupane14,borisenko14,trchang17,CaAuAs,Ag2BiO3,chang17a} and Weyl fermions\cite{xu15a,huang15,lv15,xu15b,fang12,bahadur12} familiar in the standard model of high-energy physics. Unlike high-energy physics, however, where particles are subject to the constraints of Poincar\'e symmetry, the fermions in condensed matter physics are less constrained in that they only need to respect the crystalline space-group symmetries. This easing of symmetry can lead to new fermionic particles at three-, six-, and eight-fold degenerate points in semimetals that have no high-energy counterparts\cite{wieder16,bradlyn16,chang17c,tang17,sante17,zhu16,chang17b}. In particular, three-fold fermions have been predicted in materials with triply-degenerate nodal-points (TPs) in the electronic spectrum. The triple-point semimetals (TPSs) exhibit unique topological properties such as a helical anomaly, large negative magnetoresistance, and unconventional quantum Hall effects\cite{zhu16,chang17b,Weng_TaN,Weng_WC,He_magneto,Xu_UQHE}.

The TPSs have been predicted in tungsten carbide(WC), molybdenum phosphide (MoP), antiferromagnetic half-Heusler compounds, NaCu$_3$Te$_2$, among other materials\cite{bradlyn16,zhu16,chang17b,Weng_TaN,Weng_WC,He_magneto,Xu_UQHE,yu17,xia17,wang17,yang17,gao18,huang18,kim18,song18,xiao18}. The experimental realization has been reported recently in WC\cite{ma18} and MoP\cite{lv17}. 
Notably, the TPs in a lattice can be protected either via non-symmorphic crystal symmetries \cite{bradlyn16} or through a combination of symmorphic-rotation and mirror-plane symmetries \cite{zhu16,chang17b}. Specifically, the existence of TPs in materials with symmorphic symmetries requires an inverted band-crossing of a doubly-degenerate band [two-dimensional irreducible representation (2D IR)] and a singly-degenerate band (1D IR) at a $k$-point along the rotation axis. 
Crystalline space groups containing a $C_{3v}$ ($C_{6v}$) point-group generated by a $C_3$ ($C_6$) rotation and a vertical mirror plane $\sigma_v$ supporting both 1D and 2D IRs can thus harbor these nodal-point crossings on a high-symmetry axis. 

\begin{figure}[t!]
\includegraphics[width=0.99\linewidth]{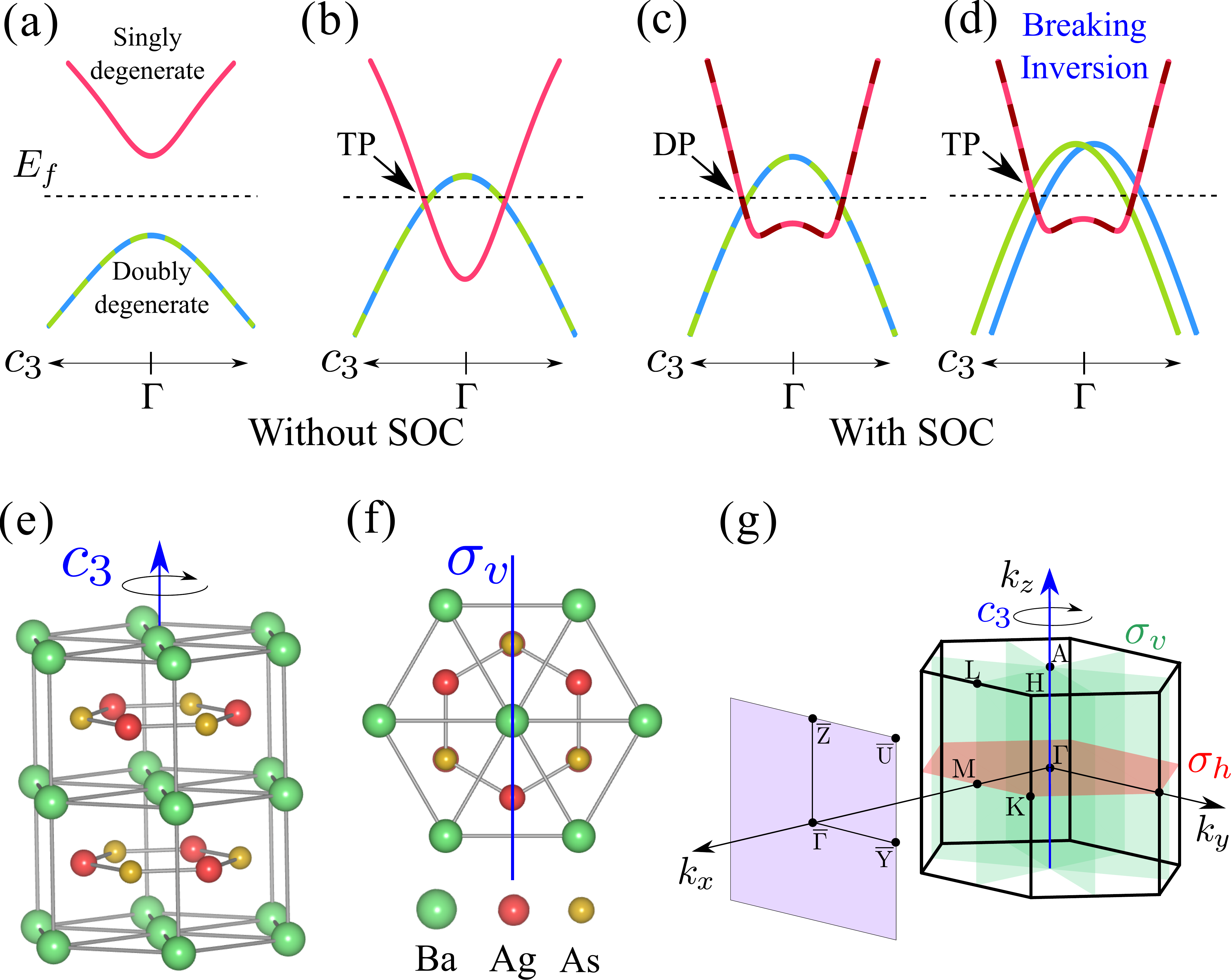}
\caption{(a)-(d) Schematic evolution of the band structure along the $C_{3}$ rotation axis in the presence of $C_{6v}$ and {time-reversal symmetries}. The symmetry-allowed singly-degenerate and doubly-degenerate bands are marked. Absence of band inversion between the singly- and doubly-degenerate bands leads to a normal insulator state in (a). 
{When there is a band inversion between these two states, a spinless TPS state appears when SOC effects are ignored, see (b). Upon turning on the SOC, we either realize (c) a stable Dirac semimetal or (d) a TPS, depending on whether or not the inversion symmetry is preserved.} 
(e) Hexagonal crystal structure of BaAgAs and its $C_{3z}$ rotation axis. Stuffed honeycomb lattice of Ag and As is also shown.  (f) Top view of the BaAgAs crystal along with its $M_{100}$ vertical mirror plane. (g) Bulk BZ and its projection onto the (100) surface plane. The horizontal mirror plane $M_{001}$ and the three symmetry equivalent vertical mirror planes $M_{100}$, $M_{010}$, and $M_{110}$ are shown with red and green colored shading.}
\label{fig:CS_BS}
\end{figure}

In Figs.~\ref{fig:CS_BS}(a)-(d), we schematically show the emergence of TPs in the Brillouin zone (BZ) in the presence of $C_{6v}$ ($C_2\otimes C_{3v}$) symmetry. In general, the doubly-degenerate and singly-degenerate bands can either stay apart (weak electron-hopping case) or cross at discrete points (strong electron-hopping case) on the $C_3$ rotation axis. The former band ordering leads to an insulating state, while the latter case gives rise to a band-inversion topological semimetal. 
In the absence of spin-orbit coupling (SOC), the nodal-crossings are stable against band repulsion due to different IRs involved and thus realize spinless TPs as shown in Fig.~\ref{fig:CS_BS}(b). When the SOC is turned on, the nodal crossings become protected by the double point-group representations at $k$ points along the $C_3$ axis, and yield a stable Dirac semimetal [Fig. 1(c)] or a TPS [Fig 1(d)] depending on whether the inversion symmetry is present or absent.
Although three fold-fermions can exist in materials with $C_{3v}$ symmetry, their realization in experimentally synthesized materials has proven challenging due to the lack of available materials in which clean triple-point crossings lie around the Fermi level.

Here, based on our first-principles calculations combined with a $\it\mathbf{k.p}$ model Hamiltonian analysis, we discuss how a gradual reduction of symmetry in Dirac semimetals along the $C_{3v}$ symmetry line can provide an effective approach for generating clean triple-point semimetals. In particular, we show that silver-based ternary BaAgAs is an ideal candidate for the experimental realization of this state of matter. BaAgAs with $C_{6v}$ symmetry is shown to host a pair of spinless TPs along its $C_{3}$ rotation axis ($z$ axis in the BZ) if the SOC is ignored. Turning on the SOC is found to transform BaAgAs into a nearly-ideal Dirac semimetal with a pair of Dirac points (DPs), which are protected by $C_3$ symmetry in a time-reversal and inversion symmetric environment. We show that when the inversion symmetry is broken via alloying in BaAgAs$_{1-x}$P$_{x}$, the material assumes a clean three-fold fermion semimetal state. We also explore the presence of a series of topological phase transition (TPTs), which are driven through the selective breaking of symmetries. Our study demonstrates that BaAgAs materials could provide an interesting platform for investigating how a variety of topological states evolve when the underlying symmetries are relaxed systematically. 

\textcolor{blue}{$ {\it Crystal~structure~and~methodology}.$\textemdash}
We perform electronic structure calculations within the framework of the density functional theory (DFT)\cite{kohan_dft} with the projector augmented wave (PAW) pseudopotentials~\cite{paw}, using the Vienna \textit{ab initio} simulation package (VASP)~\cite{kresse96, paw}. The exchange-correlation effects are treated within the generalized gradient approximation (GGA)~\cite{gga96}. SOC effects are treated self-consistently to incorporate relativistic effects. Starting with the experimental structure, we have fully optimized both the ionic positions and the lattice parameters. The tight-binding model Hamiltonian was obtained using the VASP2WANNIER90 interface, while the topological properties were obtained via the WANNIERTOOLS package~\cite{wannier90,wanniertools}. 

BaAgAs, which has already been synthesized~\cite{BaAgAs_exp}, crystallizes in a hexagonal Bravais lattice with space group $D^4_{6h}$ ($P6_3/mmc$, No. 194). Its primitive unit cell consists of a shared honeycomb lattice of Ag and As atoms which are stacked along the hexagonal $z$-axis [see Figs.~\ref{fig:CS_BS} (e) and (f)]. The Ba atoms are inserted in this stacking sequence so as to maintain the inversion symmetry of the crystal. This structure is isomorphic to the point group symmetry $C_{6v}$ or equivalently $C_2\otimes C_{3v}$ containing $C_{3}$ rotational symmetry and three symmetry-related vertical mirror planes $M_{100}$,  $M_{010}$, and $M_{110}$ and a horizontal mirror plane $M_{001}$. The bulk BZ and the associated projected (100) surface BZ are shown in Fig.~\ref{fig:CS_BS}(g). 

\begin{figure}[t]
\includegraphics[width=0.99 \linewidth]{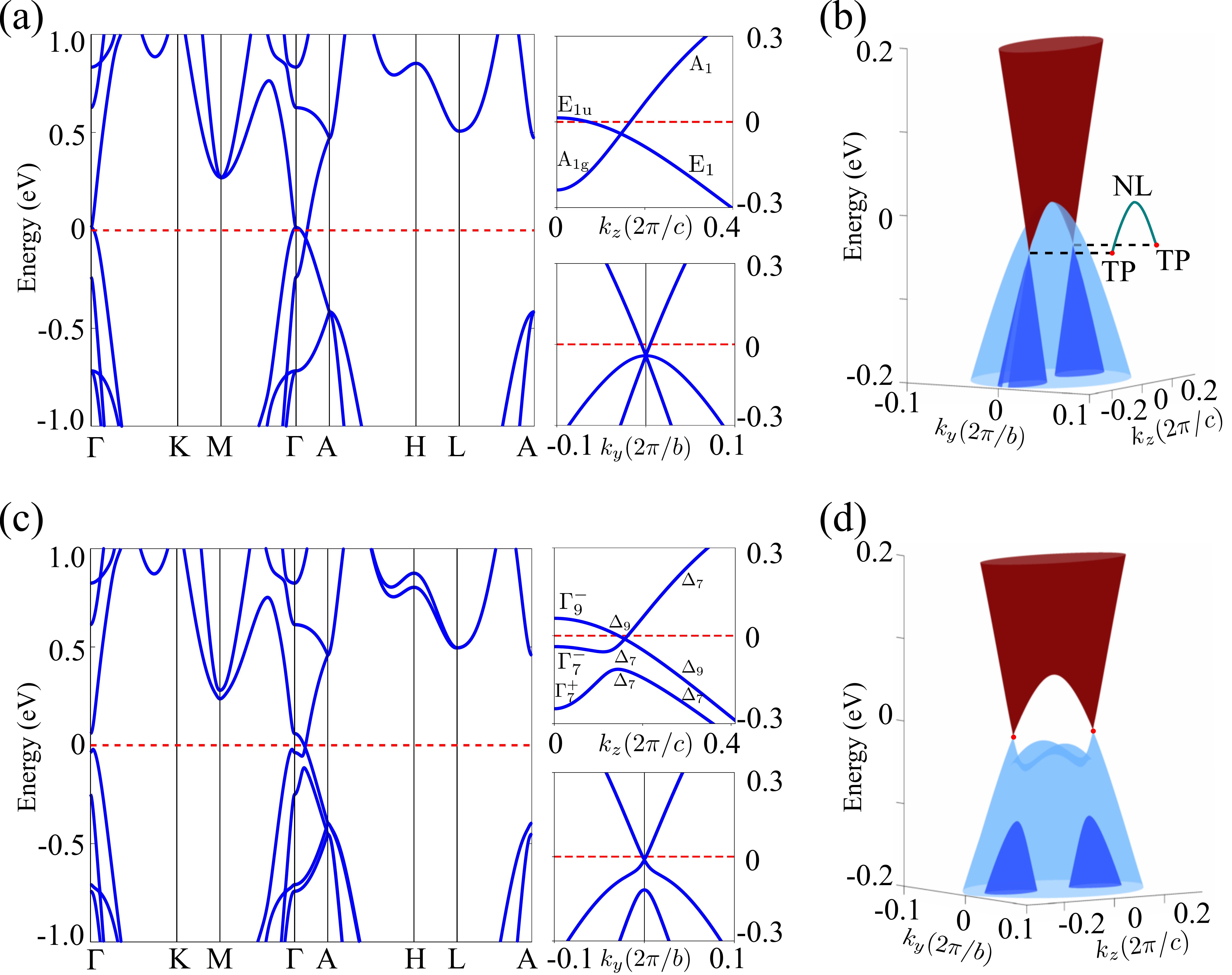}
\caption{(a) Bulk band structure of BaAgAs (without SOC). The triply degenerate nodal crossings can be seen along the $\Gamma$-$A$ direction ($C_{3}$ axis) in the BZ. The right-hand side figures show an enlarged view of the band structure along the $\Gamma$-$A$ direction (the top figures) and the in-plane direction near the TP (the bottom figures), along with some of the related irreducible representations (without SOC). (b) A 3D $E$-$k_y$-$k_z$ rendition of the bulk bands. The two symmetry-related spinless TPs and connecting single NL (green line) are shown. (c) and (d) are the same as (a) and (b) except that the SOC is included in the computations. SOC is seen to drive BaAgAs into a nearly-ideal Dirac semimetal with a pair of DPs on the $\Gamma$-$A$ line ($C_{3}$ axis) in the BZ.}
\label{fig:TP_DP}
\end{figure}

\textcolor{blue}{$ {\it Dirac~semimetal~state}.$\textemdash}
The inverted band structure of BaAgAs (without SOC) is shown in Fig.~\ref{fig:TP_DP}(a). It is seen to be a semimetal in which the doubly-degenerate $E_{1}$ and the singly-degenerate $A_{1}$ band cross along the $C_{6v}$ symmetry axis ($\Gamma$-$A$ direction). There are no other trivial band-crossings near the Fermi level. Our orbital-resolved band structure shows that the $E_{1}$ band consists of As $p$ states, while the $A_{1}$ band is mainly composed of Ag $s$ states. Since these two bands belong to different IRs of the $C_{6v}$ symmetry group along the $\Gamma$-$A$ line, their crossing point is stable against band-repulsion and realizes a TP (ignoring spin). A careful inspection of the band-crossings in the full BZ reveals that the crossing points persist only at two discrete $k$-points along the $\Gamma$-$A$ direction at $(0,0, \pm 0.146 \frac{2\pi}{c})$. Away from the $\Gamma$-$A$ line, the doubly-degenerate $E_{1}$ band splits into two as shown in the bottom inset of Fig.~\ref{fig:TP_DP}(a). Notably, a TPS can be classified based on the nodal-line connection between a pair of TPs. These nodal points in a type-A TPS are connected by one nodal line (NL) whereas they are connected by four NLs in a type-B TPS \cite{zhu16,chang17b}. Owing to the presence of the additional horizontal mirror plane $M_{001}$, BaAgAs realizes type-A TPS. A 3D illustration of the type-A band crossing is shown in Fig.~\ref{fig:TP_DP}(b) where the NL and two TPs are explicitly marked. 

The band structure of BaAgAs (with SOC) is shown in Fig.~\ref{fig:TP_DP}(c). Due to the presence of both the time-reversal and inversion symmetry, each band possesses Kramers degeneracy. Under the effects of the SOC, the doubly-degenerate $E_{1u}$ states at $\Gamma$ split into $\Gamma_9^-$ and $\Gamma_7^-$, while the singly-degenerate $A_{1g}$ state changes into the $\Gamma_7^+$ state of the double group $D^4_{6h}$. Along the $\Gamma$-$A$ direction, the symmetry is reduced to $C_{6v}$, so that the $\Gamma_7^{\pm}$ and $\Gamma_9^-$ states transform as $\Delta_{7}$ and $\Delta_{9}$, respectively. Owing to their different $C_{3}$ rotational eigenvalues, the $\Delta_{7}$ and $\Delta_{9}$ states can cross without hybridization. This leads to a stable four-fold crossing point along the $\Gamma$-$A$ direction, see Fig.~\ref{fig:TP_DP}(c). BaAgAs thus realizes a clean Dirac semimetal state with a pair of DPs on the $C_{3}$ rotation axis ($\Gamma$-$A$ line) at $(0,0, \pm 0.159~\frac{2\pi}{c})$ near the Fermi level. The energy dispersion in the vicinity of DPs is linear as illustrated in Figs.~\ref{fig:TP_DP}(c) and \ref{fig:TP_DP}(d). Keeping in mind the possible underestimation of the bandgap in the GGA, we have further confirmed our GGA-based Dirac semimetal state via parallel computations using the hybrid exchange-correlation functional. Since both the $k_z=0$ and $\pi/c$ plane are gapped, the $\mathbb{Z}_2$ topological invariants over these planes are well defined. Based on our parity analysis, we find that the $\mathbb{Z}_2$ invariant is 1 (0) for the $k_z=0 ~ (\pi/c)$ plane. The Dirac semimetal state is thus topologically protected similar to the well-known Dirac semimetal Na$_3$Bi~\cite{liu14a,wang12}.     

\begin{figure}[t]
\includegraphics[width=0.99 \linewidth]{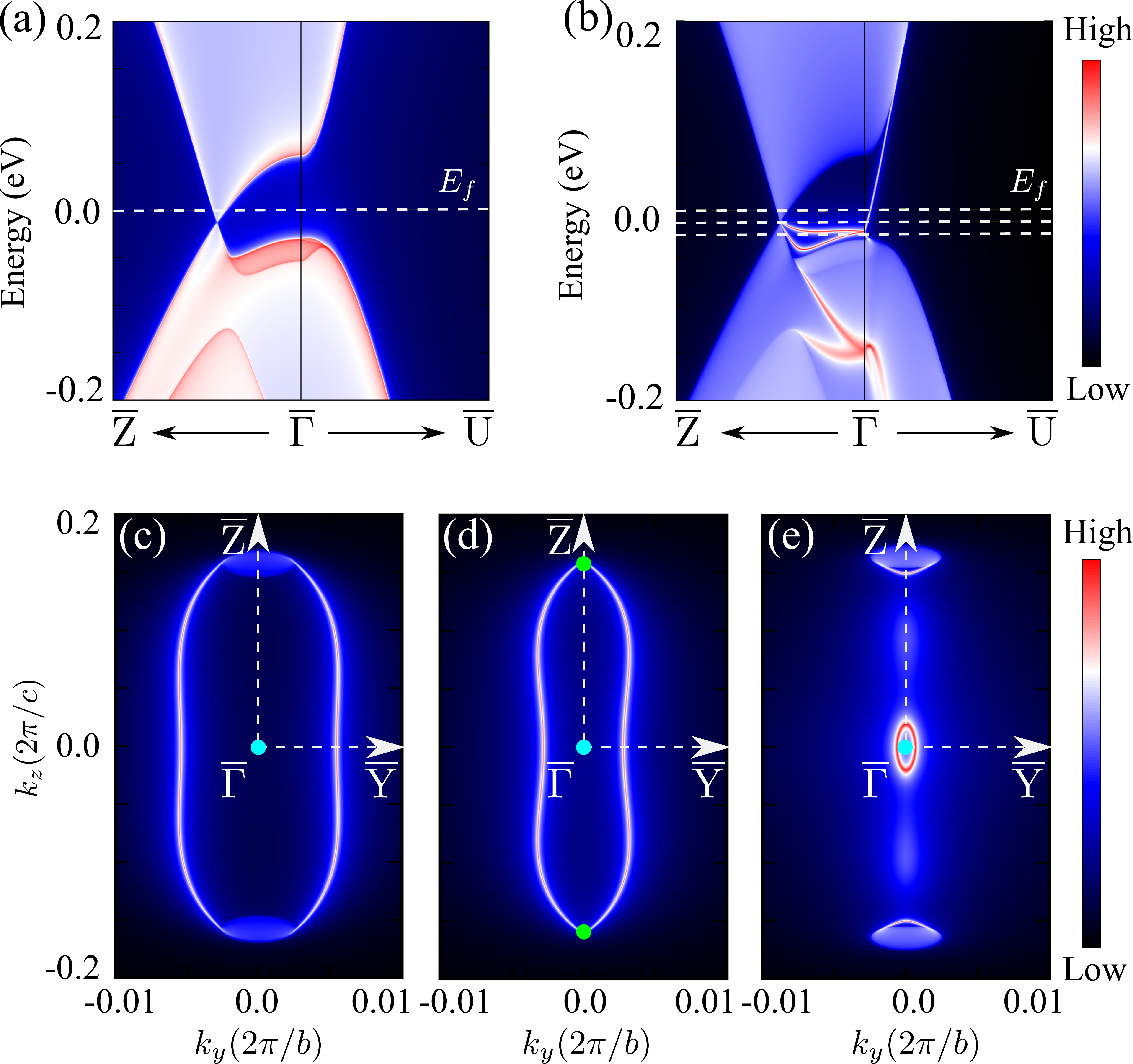}
\caption{(a) Projected bulk and (b) surface bands onto the (100) surface (with SOC). The topological surface Dirac cone with its tail connected to the projected bulk DP is visible. Constant-energy contours obtained at (c) $E= E_f$, (d) $E=E_D$, and (e) $E=E_D-12$~meV. [These energy levels are marked in (b) with dashed lines.] The DP lies at energy $E_D = -12$~meV below the Fermi level.}
\label{fig:SS}
\end{figure}

The bulk-boundary correspondence requires the presence of topological surface states connecting the bulk valence and conduction bands. To showcase this important feature, we present the (100) surface band structure of BaAgAs in Fig.~\ref{fig:SS}. The bulk-band projection on this surface is shown in Fig.~\ref{fig:SS}(a) where the DP is isolated from the trivial bulk bands along $\bar{\Gamma}$-$\bar{Z}$. A single surface Dirac cone at the $\bar \Gamma$ point is seen connecting the bulk valence and conduction bands in Fig.~\ref{fig:SS}(b). Owing to a vanishing bulk energy gap, the topological surface state terminates at the bulk DP along $\bar{\Gamma}$-$\bar{Z}$, suggesting the presence of connecting surface Fermi arcs. Figures \ref{fig:SS}(c)-\ref{fig:SS}(e) show the surface constant-energy contours at selected energy levels near the DPs. The closed surface Fermi arcs are clearly visible in Fig.~\ref{fig:SS}(d). They emanate from one projected bulk DP along $\bar{\Gamma}$-$\bar{Z}$ and terminate on the other projected DP. The evolution of these Fermi arcs above and below the DP energy is presented in Figs.~\ref{fig:SS}(c) and \ref{fig:SS}(e). 

\begin{figure}[t!]
\includegraphics[width=0.99 \linewidth]{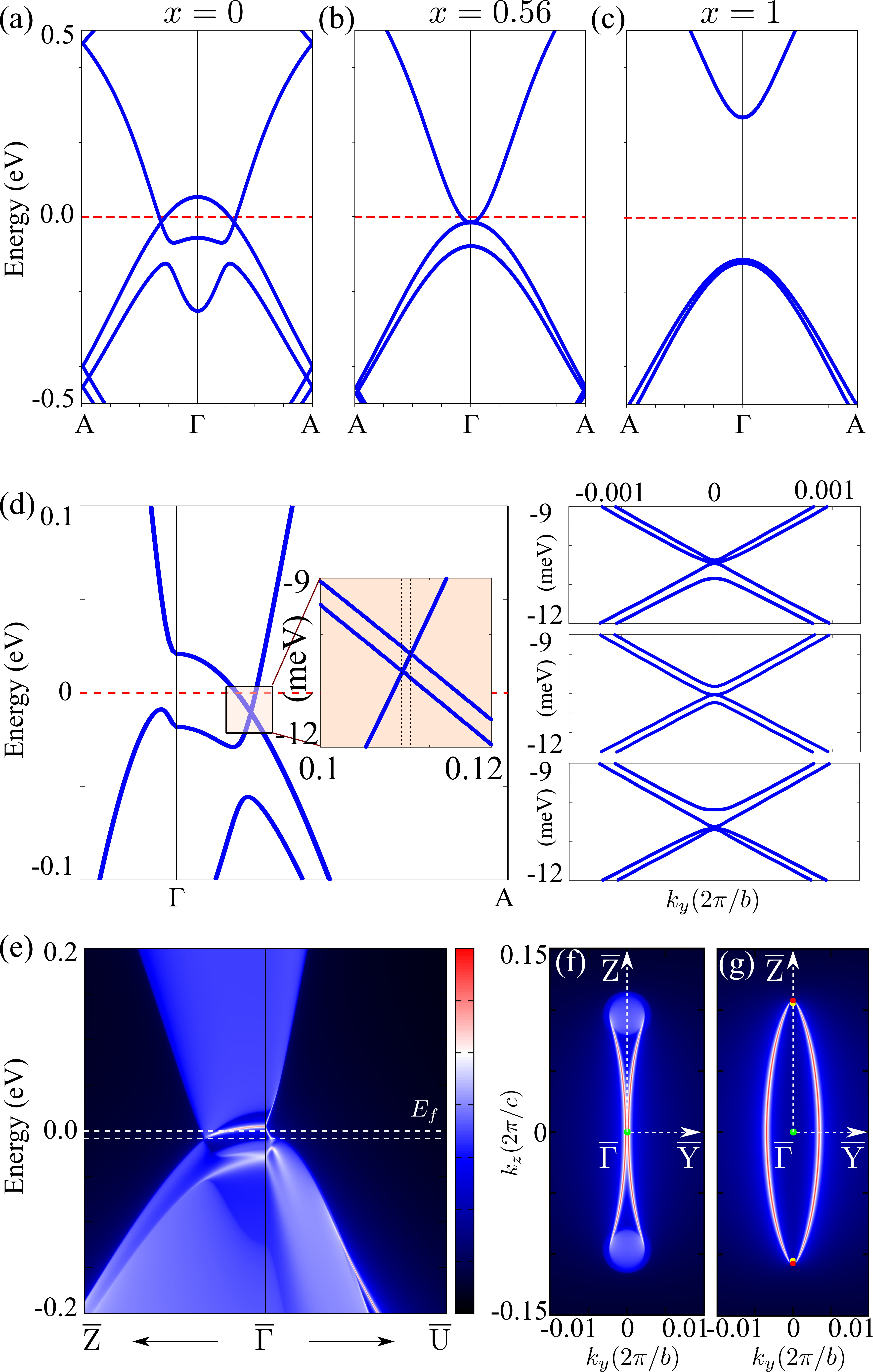}
\caption{Bulk band structure of BaAgAs$_{1-x}$P$_{x}$ alloys for different $x$ values: (a) $x= 0.0$, (b) $x= 0.56$, and (c) $x= 1.0$. The left and right hand panels show the band structure of the end compounds BaAgAs ($x= 0.0$) and BaAgP ($x= 1.0$). (d) Bulk band structure of BaAgAs$_{1-x}$P$_{x}$ supercell at $x=0.5$, with broken inversion symmetry. The inset shows a zoom in of the bands around the TP. The three right-hand panels give in-plane energy dispersion along the three line-cuts marked in the inset of (d). The top and bottom plots show the triply-degenerate band crossings. (e) Band structure of (100) surface of BaAgAs$_{0.5}$P$_{0.5}$ supercell. Constant energy contours calculated at (f) $E= E_f$ and (g) $E = -9.4$ meV. The Fermi arcs associated with the TPs are shown in (g).}
\label{fig:TPT}
\end{figure}

\textcolor{blue}{$ {\it Triple~point~semimetal~state}.$\textemdash}
We now turn to discuss the evolution of the electronic structure in BaAgAs$_{1-x}$P$_{x}$ alloys by varying the composition in order to explore possible TPT's in solid-state solutions as is the case in TlBi(Se,S)$_2$ alloys~\cite{xu11b,sato11,bahadur12}. We find that BaAgP, which is isostructural to BaAgAs, is a trivial band insulator [Fig.~\ref{fig:TPT}(c)]. The virtual crystal approximation (VCA) is used to model the electronic structure of solid solutions \cite{AB1,AB11,AB2}. In Figs.~\ref{fig:TPT}(a)-\ref{fig:TPT}(c), we show the band structures of BaAgAs$_{1-x}$P$_{x}$ for various P concentrations $x$. When $x=0.0$, there is a clear inversion between the $\Gamma_7^+$ and $\Gamma_9^-$ bands and the system realizes a topological Dirac semimetal state. As we increase $x$ to 0.56, the two bands cross at the $\Gamma$ point. With a further increase in concentration, at $x$=1.0, the band inversion vanishes and the system transitions to a trivial insulator.

The preceding results show that by varying P concentration, one can tune the topological semimetal state. However, the VCA only provides a rather simple description of the electronic structure of a solid solution. Although a more sophisticated treatment of disorder effects in alloys is possible \cite{AB3,AB4}, we have considered a supercell-based computation for the $x=0.5$ stoichiometric compound to gain some insight into the robustness of our VCA-based results. In Figs.~\ref{fig:TPT}(d)-(g), we show the results for a BaAgAs$_{0.5}$P$_{0.5}$ alloy in which the inversion symmetry of the crystal is broken, so that the symmetry of the crystal reduces from $C_{6v}$ to $C_{3v}$ on the $\Gamma$-$A$ line. As a result, the DP on the $\Gamma$-$A$ line splits into a pair of TPs [see Fig.~\ref{fig:TPT}(d)], yielding a clean TPS. This aspect is highlighted in the right-hand panels of Fig.~\ref{fig:TPT}(d) which show how the bands involved in the triply-degenerate band crossing disperse away from the $\Gamma$-$A$ line. Effects of reducing other crystalline symmetries on the emergence of topological states are further discussed in Supplemental Materials (SMs) \footnote{See Supplemental Material at http://link.aps.org/supplemental/
10.1103/PhysRevMaterials.3.071201 for further details.}. Figures \ref{fig:TPT}(e)-(g) present the topological surface states and the constant energy contours associated with the TPS state, which evolves from the topological states in the Dirac semimetal phase and connect the 
TPs. 
\footnote{{Note that the mechanism for the emergence of TPs in BaAgAs$_{0.5}$P$_{0.5}$ is similar to that proposed in  Refs.~\cite{zhu16} and \cite{chang17b}}.}
\footnote{{The TPs in  BaAgAs$_{0.5}$P$_{0.5}$ are connected by a doubly-degenerate nodal line in sharp contrast to the case of a non-symmorphic TPS \cite{bradlyn16}, and support unique Fermi surfaces which essentially touch at the doubly-degenerate band. These features are distinct from those of other known semimetals and can lead, for example, to magnetic tunneling/breakdown in quantum oscillations.}}

\textcolor{blue}{$ {\it Model~Hamiltonian}.$\textemdash}
In order to better understand the emergence of the TPS state in BaAgAs alloys, we now discuss a viable low-energy $\mathbf{\mathbf{k.p}}$ model Hamiltonian using the theory of invariants. Using the $\Delta_{9}$ and $\Delta_{7}$ states of the $C_{6v}$ group as the basis with components $\Delta_9(+3/2)$, $\Delta_9(-3/2)$, $\Delta_7(+1/2)$, and  $\Delta_7(-1/2)$, the minimal four-band Hamiltonian around a DP to $\mathcal{O}(k_{x,y}^2)$ and $\mathcal{O}(k_z)$ is given by,
\begin{equation}\label{eq:1}
{\cal H}_{\rm eff}(\mathbf{k}) = \left( 
\scalemath{0.8}{
\begin{array}{cccc}
\varepsilon _{1}(\mathbf{k}) & 0 & c(\mathbf{k}) & d(\mathbf{k}) \\ 
0 & \varepsilon _{1}(\mathbf{k}) & -d^{\ast }(\mathbf{k}) & c^{\ast }(%
\mathbf{k}) \\ 
c^{\ast }(\mathbf{k}) & -d(\mathbf{k}) & \varepsilon _{2}(\mathbf{k}) & 0 \\ 
d^{\ast }(\mathbf{k}) & c(\mathbf{k}) & 0 & \varepsilon _{2}(\mathbf{k})%
\end{array}}%
\right)
\end{equation}
where $\varepsilon _{1}(\mathbf{k}) =A_{1}k_{\perp }^{2}+B_{1}k_{z}$, $\varepsilon _{2}(\mathbf{k}) =A_{2}k_{\perp }^{2}+B_{2}k_{z}$, $c(\mathbf{k}) =C(k_{x}+ik_{y})^{2}$, and $d(\mathbf{k}) =\left( D+D^{\prime }k_{z}\right) (k_{x}+ik_{y})$ using $k_{\perp}^{2}=k_{x}^{2}+k_{y}^{2}$. The eigenenergies of the Hamiltonian ${\cal H}_{\rm eff}(\mathbf{k})$ are
\begin{eqnarray}\nonumber
E(\mathbf{k})&=&A_{+}k_{\perp }^{2}+B_{+}k_{z}\\ \nonumber
&\pm& \sqrt{\left[ A_{-}k_{\perp}^{2}+B_{-}k_{z}\right] ^{2}+C^{2}k_{\perp }^{4}+\left( D+D^{\prime}k_{z}\right) ^{2}k_{\perp }^{2}}, 
\end{eqnarray}
with $A_{\pm }=\frac{A_{1}\pm A_{2}}{2}$ and $B_{\pm }=\frac{B_{1}\pm B_{2}}{2}$, from which we identify a DP at $\mathbf{k}=\mathbf{0}$.
The material-dependent parameters in the Hamiltonian obtained numerically by fit to our first-principles results are as follows:   
$A_1$ = -17.78 eV$\mathrm{\AA}^2$,  $A_2$ = 77.26 eV$\mathrm{\AA}^2$, $B_1$ = 1.67 eV$\mathrm{\AA}$, $B_2$ = -0.85 eV$\mathrm{\AA}$, $C$ = -0.08 eV$\mathrm{\AA}^2$, $D$ = 4.11 eV$\mathrm{\AA}$, and  $D^{\prime}$ = 0.75 eV$\mathrm{\AA}^2$. The TPS state in BaAgAs$_{1-x}$P$_{x}$ alloy is described directly by breaking inversion symmetry through adding $\Delta {\cal H} =  h \left( 
\begin{array}{cc}
1 & 0 \\ 0 & 0
\end{array}
 \right) \otimes \sigma_y$, with $h=-0.00022$ eV to  ${\cal H}_{\rm eff}(\mathbf{k})$ in Eq. (\ref{eq:1}). The DP now splits into a pair of TPs located at $k_{TP} = (0,0,\pm \frac{h}{B_1-B2})$. It should be noted that there are many forms of perturbation that can break inversion symmetry. Here, we do so by breaking $C_{2}$ that reduces $C_{6v}$ symmetry to $C_{3v}$ along the $\Gamma$-$A$ line. 

\textcolor{blue}{$ {\it Summary~ and~conclusions}.$\textemdash}
The distinct topological states that we have identified in the BaAgAs family suggest that these materials would provide a robust experimental platform for exploring a variety of different topological phases in a single materials family. Notably, BaAgAs and BaAgP have been synthesized. These end compounds assume different topological phases, but are isostructural, which should make them amenable to alloying to provide a tunable material, much as the case of the topological insulator TlBi(Se,S)$_2$. Our predicted topological states lie close to the Fermi level without the background of any trivial bands. These observations indicate that the theoretically predicted topological band features would be accessible to angle-resolved-photoemission spectroscopy (ARPES) and scanning-tunneling-spectroscopy (STS) experiments.   

In summary, based on our first-principles calculations combined with a $\mathbf{k.p}$ model Hamiltonian analysis, we show that the BaAgAs family can realize a tunable topological semimetal state via a gradual lowering of the crystal symmetry. In the absence of SOC effects, BaAgAs harbors a clean TPS state with a pair of $C_{6v}$ symmetry-protected TPs. Inclusion of the SOC drives the material into a nearly-ideal topological Dirac semimetal with a pair of DPs near the Fermi level along the $C_{3}$ rotation axis. We show that  when the inversion symmetry is relaxed in BaAgAs$_{1-x}$P$_{x}$ alloys, one can obtain a clean TPS at $x=0.5$. Our study reveals that BaAgAs materials family offers an exciting platform for exploring tunable topological semimetal states and the unique physics associated with these states. 

$Acknowledgements $\textemdash
The work at Shenzhen University was supported by the Shenzhen Peacock Plan (KQTD2016053112042971) and the Science and Technology Planning Project of Guangdong Province (2016B050501005). The work at Northeastern University was supported by the U.S. Department of Energy (DOE), Office of Science, Basic Energy Sciences Grant No. DE-FG02-07ER46352, and benefited from Northeastern University’s Advanced Scientific Computation Center and the National Energy Research Scientific Computing Center through DOE Grant No. DE-AC02-05CH11231. S.M.H. is supported by the Ministry of Science and Technology (MOST) in Taiwan under Grant No. 105-2112-M-110-014-MY3. T.-R.C. was supported from Young Scholar Fellowship Program by Ministry of Science and Technology (MOST) in Taiwan, under MOST Grant for the Columbus Program MOST108-2636-M-006-002, National Cheng Kung University, National Center for Theoretical Sciences (NCTS), and Higher Education Sprout Project, Ministry of Education to the Headquarters of University Advancement at National Cheng Kung University. This work is supported partially by the MOST, Taiwan, Grants No. MOST 107-2627-E-006-001.
%
H. L. acknowledges Academia Sinica, Taiwan for support under Innovative Materials and Analysis Technology Exploration (AS-iMATE-107-11). The work at IIT Kanpur benefited from the high-performance facilities of the computer center of IIT Kanpur. 

\bibliographystyle{prsty}
\bibliography{BaAgAs}
\end{document}